\documentclass[a4paper]{article}
\usepackage{amsmath, amssymb}
\usepackage{bm}
\usepackage[dvipsnames]{xcolor}
\usepackage{cite}
\newcommand{\be}{\begin{equation}}
\newcommand{\ee}{\end{equation}}
\newcommand{\bea}{\begin{eqnarray}}
\newcommand{\eea}{\end{eqnarray}}

\usepackage{tikz}
	\usetikzlibrary{calc}
	\usetikzlibrary{arrows.meta}
	\usetikzlibrary{decorations.markings}

\usepackage[
	hidelinks,
    colorlinks,
    citecolor = Green,
    linkcolor = blue,
    urlcolor  = magenta
]{hyperref}

\usepackage[
    left   = 20mm,
    right  = 20mm,
    top    = 20mm,
    bottom = 30mm
]{geometry}

\begin{document}

\title{Critical Phases of the extended isotropic $XX$ chain with four-spin interaction}

\author{Nika Kurdadze,$^{a}$
\, Giorgi Gogaberishvili,$^{a}$ 
 and G.I. Japaridze$^{a,b}$\footnote{\text{e-mail: gia$_{-}$japaridze@iliauni.edu.ge
 }}}
\date{
\small
$^{a}$\textit{Ilia State University, Cholokashvili Avenue 3-5, 0162 Tbilisi, Georgia}\\
$^{b}$\textit{Andronikashvili Institute of Physics, Tamarashvili str.~6, 0177 Tbilisi, Georgia}\\[2ex]
\today
}

\maketitle

\begin{abstract}

Using the Jordan-Wigner transformation we calculate exactly the ground state  and low-temperature thermodynamic properties of the spin $S=1/2$ isotropic $XX$ chain with four spin interaction. In terms of the equivalent spinless fermion (SF) representation the system is viewed as a lattice fermion gas with nearest-neighbor ($J$) and next-next-next-neighbor ($J^{\ast}/4$) hopping. 
It is shown that with the increase of four spin coupling, at $J^{\ast}_{c} = 4J/3$ the system experiences the Lifshitz type topological phase transition characterized by the tripling of Fermi points. The quantum phase transition (QPT) point marks transition from a gapless spin-liquid phase of standard $XX$ chain into again a gapless spin-liquid phase with different character of power-law decay of spin correlations. At the transition point the free fermion dispersion relation shows flattering at Fermi points, what determines singular character of density of states $\rho(\omega)\sim (\omega/J)^{-2/3}$ and as a consequence unconventional temperature dependence of heat capacity of the system $C\sim (T/J^{\ast})^{1/3}$, and singular magnetic susceptibility of the system $\chi(H)\sim (H/J^{\ast})^{-2/3}$.

In the case of alternating magnetic field the system is characterized by the rich ground state phase diagram  which contains fully polarized (ferromagnetic), gapped antiferromagnetic (AFM) and spin liquid phases. At the transition point from the gapped AFM phase into the gapless polarized spin liquid phase the system shows rapid increase of magnetization 
$m\sim(H-H_c)^{1/6}$ and  magnetic susceptibility a singular behavior as $\chi(H)\sim (H-H_c)^{-5/6}$.

\end{abstract}

\hspace{2cm} PACS:  {75.10.Pq -Spin chain models, 75.40.-s Critical-point effects,

\hspace{3.1cm} 75.10.Jm Quantized spin models.}

\section{Introduction}
\label{intro}

The study of the quantum phase transitions (QPTs) \cite{Sachdev_Book_99} in low-dimensional systems has attracted much of current interest particularly in the context of applications in modern quantum information technologies \cite{Wen_19}. The one-dimensional quantum spin systems, because of rich quantum features and complex ground state phase diagrams,  serve as a fascinating playground to investigate QPTs. \cite{GNT_book_98,Giamarchi_book_04,Takahashi_Book_05,Cabra_Pujol_04,
Vasiliev_Volkova_18}

The idea that the lattice models with more than two-body interactions may lead to a rich variety of critical behavior found its confirmation in several two-dimensional (2D) models of statistical physics: the Ashkin-Teller model~\cite{AT_1943}, the Baxter model~\cite{Baxter_1972} and the Baxter-Wu model~\cite{Baxter_Wu_1973} involving four- and three-spin interactions, respectively. Their critical behavior differs drastically from that of Onsager's solution of the 2D Ising model. Later, several interesting models with multi-spin interaction have been proposed and studied~\cite{Gross_Mezard_84, Gardner_85, m-spin_Ising_Kolb_Penson_86, 3-spin_1988_Penson, m-spin_Ising_1998, 3-spin_Ising_1999, 3-spin_Ising_2009} mostly in the context of spin glass physics~\cite{Gross_Mezard_84,Gardner_85}, quantum statistics~\cite{m-spin_Ising_Kolb_Penson_86, 3-spin_1988_Penson, m-spin_Ising_1998, 3-spin_Ising_1999, 3-spin_Ising_2009,Velonakis_2013,Monroe_2024} and recently in the context of ladder systems with flux \cite{Garuchava_etal_25,Garuchava_26}.

A special subset of Hamiltonians with multispin interactions belongs to the class of one-dimensional (1D) exactly solvable models \cite{Tsvelik_90,Frahm_92,Muramoto_Takahashi_99,ZKZ_2001,Zvyagin_Kluemper_03,Franchini_2017}. The advantages of studying 1D exactly solvable spin models permit us to obtain an exact description of the ground-state properties and thermodynamic characteristics of the system using nonperturbative methods. A special class of spin $S=1/2$ chain models with multi-spin interactions, allowing an exact solution in terms of a {\em Fermi gas of spinless Fermions}, was proposed by M. Suzuki more than fifty years ago \cite{Suzuki_71}. Similar to the standard spin $S=1/2$ $XX$ chain \cite{LSM_61,Katsura_62}, Suzuki models can be rewritten in terms of free spinless fermions on a 1D lattice by means of the Jordan-Wigner transformation \cite{JW_1928}, but with long-distance hopping determined by the structure of the multi-spin term. As a result, Suzuki models provide a unique opportunity for realizing topological Lifshitz transitions in 1D \cite{Lifshitz_1960,Volovik_2017} where, for a given band filling, the number of Fermi points changes with change of the  multi-spin coupling. As a result, the Suzuki class of extended $XX$ models with multi-spin interactions has become a useful playground for studying topological quantum phase transitions and has attracted considerable current interest
\cite{Gottlieb_Roessle_99,TJ_2003,Lou_etal_04,Zvyagin_05,Zvyagin_2006,Derzhko_etal_08,Derzhko_etal_09,
Derzhko_etal_11a,Ribeiro_etal_11,Derzhko_etal_11b,Ohanyan_etal_15a,Ohanyan_etal_15b,Dutta_etal_16,
Wuhan_team_18,Harbin_team_21,Kaiyuan_Cao etal_2022,Mahdavifar_etal_24,Malakar_Ghosh_2024,Mahdavifar_etal_26}. Independently, topological Lifshitz transitions in one-dimensional systems have attracted particular current interest both in the context of ultracold atomic systems \cite{Cazalilla_etal_2022} and ladder systems with orbital currents in the presence of a penetrating magnetic flux \cite{BN_1_EPJB}.

In this paper we consider the extended $XX$ model with four spin coupling,  given by the  Hamiltonian 
\bea \label{GXYhamiltonian}
& H_{XzzX}  =   -J\sum_{n=1}^{N}\left(S^{x}_{n}S^{x}_{n+1} + 
S^{y}_{n}S^{y}_{n+1}\right) - J^{\ast} \sum_{n=1}^{N}  \left(S^{x}_{n}S^{x}_{n+3} +
S^{y}_{n}S^{y}_{n+3}\right)S^{z}_{n+1}S^{z}_{n+2} \, , &
\eea
where $J,J^{\ast} > 0$, $N$ is the total length of the chain, and $S^{\beta}_{n}$ ($\beta=x,y,z$) are spin-1/2 operators defined on the n-th site. 
The model (\ref{GXYhamiltonian}) attracted our attention owing to several important advantages: it is exactly solvable, and its exact solution is expressed in terms of a Fermi gas of spinless fermions (SF).

Moreover, as shown in this paper, 

\begin{itemize}
\item{with increasing four-spin coupling, at $J^{\ast}_{c} = 4J/3$ the system undergoes a quantum phase transition (QPT) between two gapless spin-liquid phases
characterized by different thermodynamic and magnetic properties;}
\item{in terms of the equivalent free SF model, the transition appears as a Lifshitz-type topological phase transition characterized by the tripling of the Fermi points;}
\item{at the transition point, the particle spectrum at the Fermi points becomes substantially flat ($E(k)\simeq J^{\ast}_{c}\cos^{3}k$), which determines the anomalous magnetic and low-temperature thermodynamic properties of the system near the transition point;}
\item{in the presence of an applied uniform and/or staggered magnetic field, the system is characterized by a rich ground-state phase diagram containing {\em ferromagnetic, antiferromagnetic and three
different spin-liquid phases}.}
\end{itemize}

In this paper, we restrict our consideration to the most interesting unconventional properties of the system near the topological Lifshitz transition. The paper is organized as follows. In the forthcoming Section, we discuss the spinless fermion representation of the model. In Section III, the ground-state quantum phase transition is considered, and anomalies in the low-temperature thermodynamics of the
system at the QPT point are discussed. In Section IV, the behavior of the system at the transition point in the presence of uniform and staggered magnetic fields is studied. Finally, the obtained results are summarized in Section V.

\section{Spinless Fermion representation}

The Hamiltonian (\ref{GXYhamiltonian}) can be transformed into spinless fermion representation by conducting a Jordan-Wigner transformation \cite{JW_1928},
which is defined as
\begin{eqnarray} \label{JWtransformation}
\nonumber\\ 
S^{+}_{n}= S^{x}_{n} + {\it i}S^{y}_{n}&=&
\prod_{m=1}^{n-1}\left(1-2c_{m}^{\dagger}c_{m}\right)c_{n}^{\dagger}\nonumber\\
S^{-}_{n}= S^{x}_{n} - {\it i}S^{y}_{n} &=& \prod_{m=1}^{n-1}c_{n}
\left(1-2c_{m}^{\dagger}c_{m}\right)\\ 
S^{z}_{n} &=& c_{n}^{\dagger}c_{n} -\frac{1}{2} \nonumber 
\end{eqnarray}
where $c_{n}^{\dagger}, c_{n}$ are the spinless fermion (SF) creation and annihilation 
operators, respectively. In fact, the Hamiltonian (\ref{GXYhamiltonian}) is transformed 
to a free spinless fermion model as
\bea \label{GXYSFhamiltonian}
{\cal H} & = & -\frac{J}{2}\sum_{n}\left(c_{n}^{\dagger}c_{n+1}+ c^{\dagger}_{n+1}c_{n}\right)-\frac{ J^{\ast}}{8}\sum_{n}\left(c_{n}^{\dagger}c_{n+3}+ c^{\dagger}_{n+3}c_{n}\right)\, . 
\eea

We can diagonalize the Hamiltonian (\ref{GXYSFhamiltonian}) by means of the Fourier 
transformation to obtain 
\begin{equation} \label{GXYSFhamiltonianEk}
{\cal H}=\sum_{k}E(k)c^{\dagger}(k)c(k)
\end{equation} 
where
\be \label{Spectrum}
E(k)=-J\left(\cos(k)+\frac{\alpha}{4}\cos3k\right)=-J\big[(1-\frac{3\alpha}{4})\cos k+\alpha\cos^{3} k\big]
\ee
and $\alpha=J^{\ast}/J$. 

\begin{figure}[!t]
\begin{center}
\includegraphics[width =0.6\columnwidth]{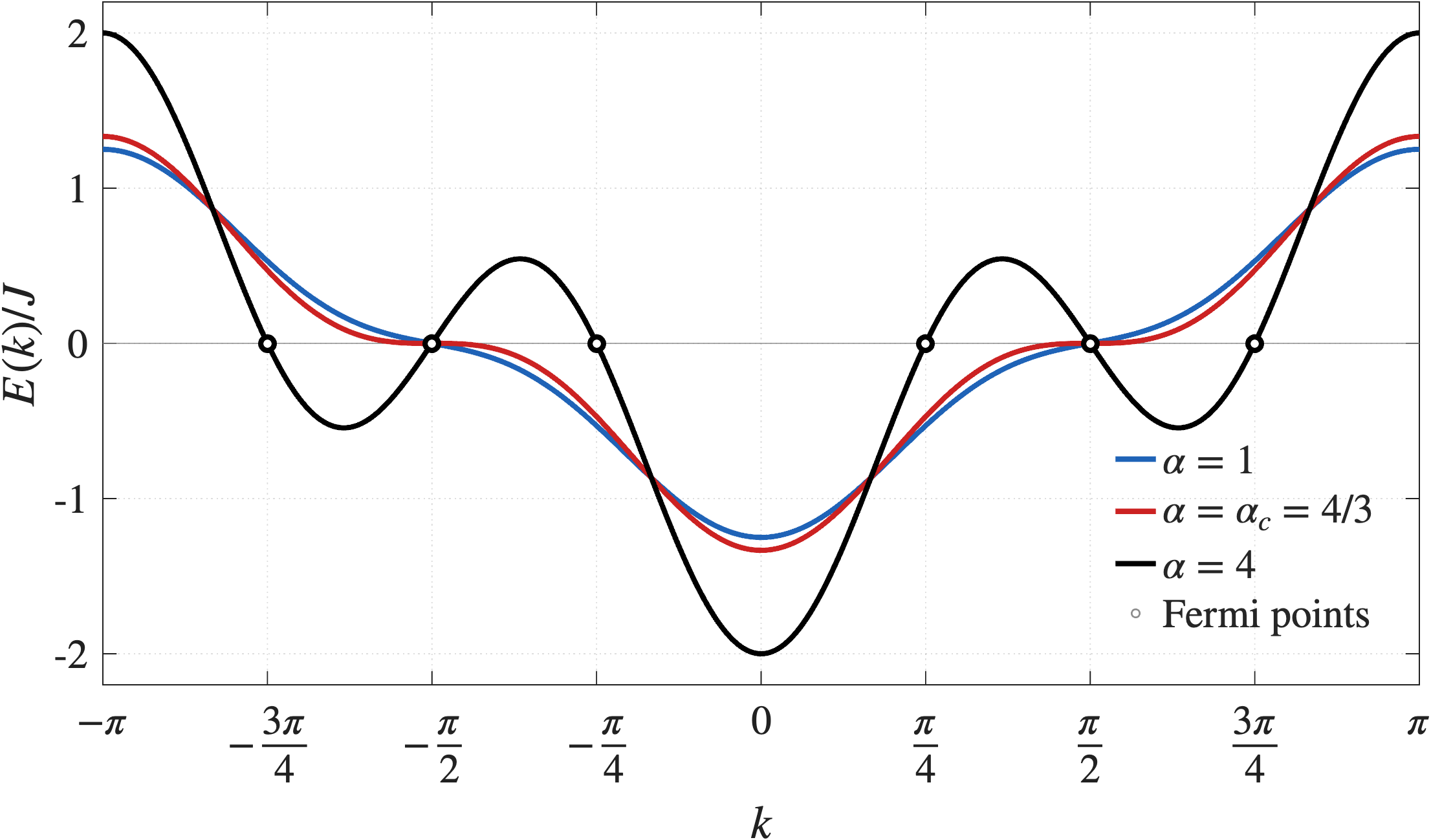}
\caption{ Dispersion relation for different positive values of the parameter $\alpha$. The blue line correspond to values of $\alpha$ below the Lifshitz transition, where the spectrum is characterized by two Fermi points. The red line corresponds to the spectrum at the critical point,
$\alpha=\alpha_{c}=4/3$. The black line corresponds to the spectrum with six Fermi points at $\alpha=4$.}
 \label{fig:Spectrum_alpha}
 \end{center} 
\end{figure}

In the thermodynamic limit ($N \rightarrow \infty$), the ground state of the system corresponds to the configuration in which all states with $E(k) \leq 0$ are filled and those with $E(k)>0$ are empty.  Because the original spin Hamiltonian conserves total spin along the z axis: [H; Sz] = 0 and is invariant with respect to inversion of “z” components of all spins $ S_{i}^{z} \rightarrow  - S_{i}^{z}$, , in absence of magnetic field the ground state in the spin-liquid phase is always nonmagnetic $M=\sum_{i} S_{i}^{z}=0$. After the Jordan-Wigner transformation, this symmetry is manifested as conservation of the total fermion number
and zero chemical potential ensuring the half-filling of the equivalent SF model in absence of magnetic field.

 The boundaries of these regions are marked by the corresponding Fermi points, which are solutions of the equation $E(k)=0$. For $\alpha>0$ and $0 \leq \alpha \leq \alpha_{c}=4/3$, the spectrum is characterized by two Fermi points $\pm k_{F}^{0}= \pi/2$, while for $\alpha > \alpha_{c}$ it is characterized by six Fermi points.(see Fig. \ref{fig:Spectrum_alpha})
\begin{equation} \label{FermiPointsAlpha}
\pm k_{F}^{0}= \pi/2, \, \quad  \pm k_{F}^{\pm}=\pm\left(\pi/2 \pm \arcsin\sqrt{3/4-1/\alpha}\right), 
\end{equation} 
\begin{equation} \label{FermiPointsAlpha_1}
\pm k_{F}^{0}= \pi/2, \, \quad  \pm k_{F}^{\pm}=\pm\left(\pi/2 \pm k_{0}\right), 
\end{equation} 
where 
$$
k_{0}=\arcsin\sqrt{3/4-1/\alpha}\, .
$$
Thus, the critical value of the parameter $\alpha=\alpha_{c}=4/3$ marks a Lifshitz-type \cite{Lifshitz_1960,Volovik_2017} topological quantum phase transition (QPT) in the ground state of the model. The absolute minimum of the spectrum, $E_{min}=-J(1+\frac{\alpha}{4})$, is reached at $k=0$, while the absolute maximum, $E_{max}=J(1+\frac{\alpha}{4})$, is reached at $k=\pi$. 

At $\alpha>\alpha_{c}$, the spectrum is characterized by additional local extrema at 
$$
k=k^{\ast}=\arccos\left(\pm\sqrt{1/4-1/(3\alpha)}\right)
$$ 
which appear as a bump and a concavity in the dispersion curve. The height and depth of these local extrema
\begin{equation} \label{FermiPointsSlphaP}
\pm E(k^{\ast})= \pm \frac{J^{\ast}}{4}\left(1-\frac{4}{3\alpha}\right)^{3/2}\, ,
\end{equation} 
increase with increasing $\alpha$, approaching the maximum and minimum values $J^{\ast}/4$ and $-J^{\ast}/4$, respectively, as $\alpha \rightarrow \infty$.  

At the critical point $\alpha=\alpha_{c}$, the spectrum in the vicinity of the Fermi points $k_{F}=\pm \pi/2$ is cubic:
\be \label{Linearized_Spectrum}
E(\pm \pi/2+p)=\pm J^{\ast}p^{3}\left[1+ {\cal O}(p^{2})\right]
\ee
for $p \ll |k_{F}^{0}|$. This determines the anomalous properties of the system at the transition point, and below we focus mainly on the properties of the system near this point.

\section{Quantum Phase Transition in the ground state}

In the ground state, all negative-energy states are filled, while the positive-energy states are empty. Accordingly, the ground state energy (GSE) of the system is given by
\begin{equation} \label{GSE_total}
{\cal E}_{0}(\alpha)=\frac{L}{2\pi}\int_{\{\Lambda\}}E(k)\,dk
\end{equation}
where the integration region is $\{\Lambda\}=[-\pi/2,\pi/2]$ for $\alpha \leq 4/3$ and  
$\{\Lambda\}=[-k^{-}_{F},k^{-}_{F}]\cup [-\pi/2,-k^{+}_{F}]\cup [\pi/2,k^{+}_{F}]
$ for $\alpha > 4/3$. Using (\ref{Spectrum}) and (\ref{FermiPointsAlpha_1}), after integration we obtain the following GSE density
\begin{equation} \label{GSE_density}
\frac{1}{L}{\cal E}_{0}(\alpha) \equiv e_{0}(\alpha)=-\frac{J}{2\pi}\cdot
\left\{ \begin{array}{l@{\quad}}2-\frac{\alpha}{6}
 \hskip3.0cm {\mbox{at}}\quad \alpha\leq 4/3  \\\frac{\alpha+4}{3}\sqrt{1+\frac{4}{\alpha}}-2+\frac{\alpha}{6}  \hskip0.7cm
{\mbox{at}}\quad \alpha > 4/3 \end{array}\right. \, .
\end{equation}
\begin{figure}[!t]
\begin{center}
  \includegraphics[width =0.7\columnwidth]{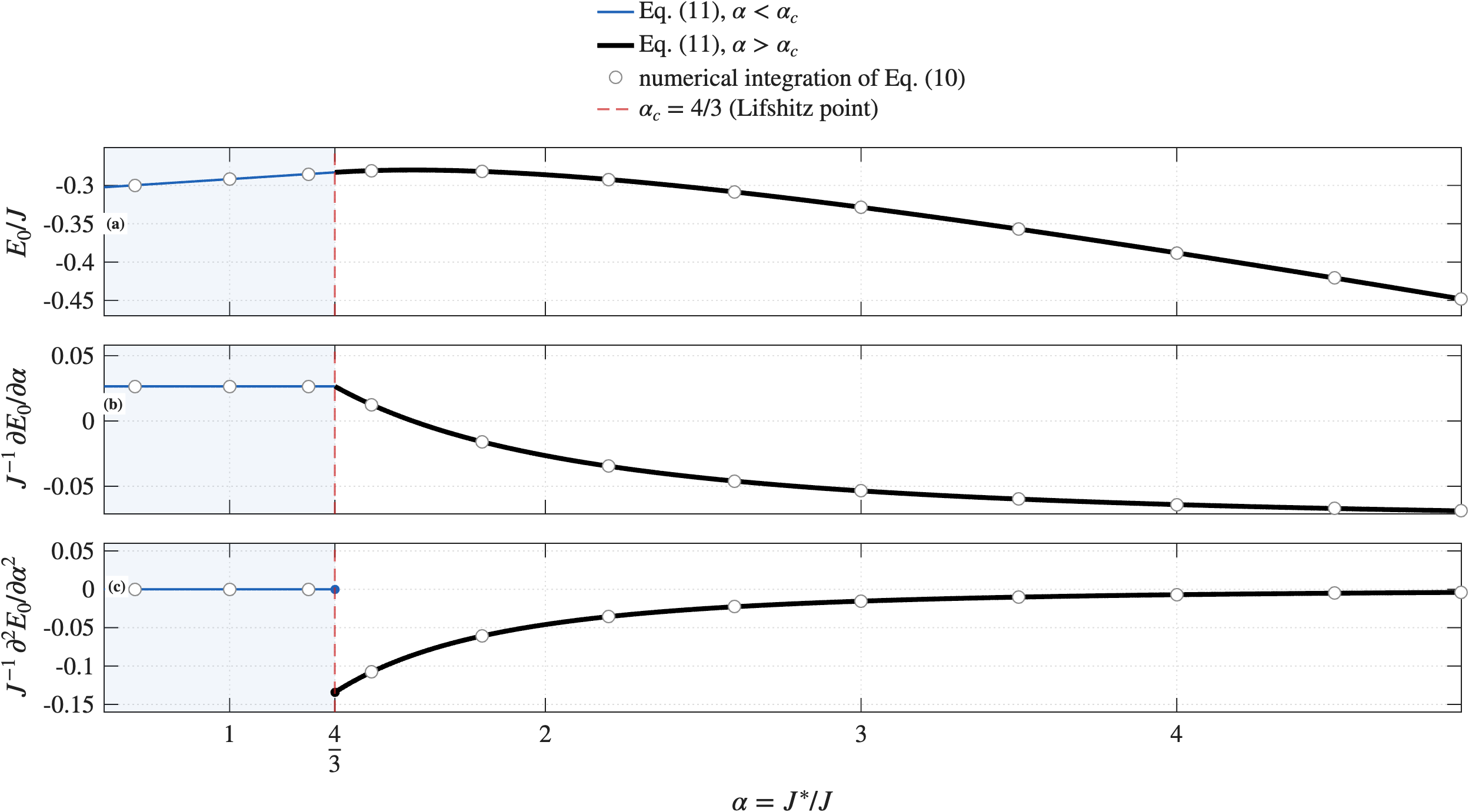}
  \caption{The ground state energy and its first and second derivatives as functions of the parameter $\alpha$.}
  \label{fig:GSE_and_derivatives}
 \end{center} 
\end{figure}
The Livshitz type phase transitions, at the transition point besides the change of the topology of Fermi surface, (of the number of Fermi points in 1D) are characterized, due to these changes,  by the anomalies in the density of states, and hence in the ground state energy (GSE) of the system \cite{Lifshitz_1960}. To address these aspect, in Fig.~\ref{fig:GSE_and_derivatives}, we plot the GSE density, as well as its first and second derivatives, as functions of the parameter $\alpha$. At $\alpha=\alpha_{c}=4/3$, we observe non-monotonic behavior of the ground state energy. In particular, the first derivative of the ground state energy with respect to the parameter $\alpha$ shows a kink, while the second derivative shows a discontinuity. We treat this discontinuity in the generalized stiffness,
$$
\xi(\alpha)=-\frac{\partial^{2}{\cal E}_{0}(\alpha)}{\partial\alpha^{2}}\, ,
$$ 
as an indicator of a second-order phase transition in the ground state of the system with increasing multi-spin coupling $J^{\ast}=\alpha J$. 

Note that, in the absence of a magnetic field, the width of the $\{\Lambda\}$ region in momentum space where $E_{k}(\alpha)\leq 0$ is equal to $\pi$ for arbitrary $\alpha$.   Therefore, in marked contrast to the extended $XzX$ chain with three-spin coupling \cite{TJ_2003}, the band of the equivalent spinless fermion model is always half-filled. Using (\ref{JWtransformation}), one easily finds that the net magnetization of the system in the ground state is
\begin{equation} \label{Magnetization_zero}
m=\frac{1}{L}\sum_{i}S_{i}^{z}=
\frac{1}{2\pi}\int_{-\pi}^{\pi}n(k)dk-\frac{1}{2}=\frac{1}{2\pi}\int_{\{\Lambda\}}dk-\frac{1}{2}=0, 
\end{equation}
where the Fermi distribution function of spinless  in the ground state $n(k)=\langle a_{k}^{\dagger}a_{k}\rangle_{0}=1$ for $k\in \{\Lambda\}$ and is zero otherwise. 

An important characteristics of the Lifshitz type topological phase transitions is the anomaly in density of states at the transition point \cite{Lifshitz_1960}. At the critical point $\alpha=\alpha_{c}=4/3$, the dispersion relation of spinless fermions becomes $\omega(k)=J^{\ast}\cos^{3}(k)$. Therefore, the density of states 
\begin{equation} \label{DOS}
\rho(\omega)=(d\omega/dk)^{-1}=\frac{1}{3J^{\ast}_{c}(\omega/J^{\ast}_{c})^{2/3}\sqrt{1-(\omega/J^{\ast}_{c})^{2/3}}}
\end{equation} 
shows a profound singular behavior $\rho(\omega)\sim (\omega/J^{\ast})^{-2/3}$ in the low-energy limit as the consequence of the flattening of the spectrum at the critical point.
\subsection{Thermodynamics}

This singularity in the density of states at the quantum critical point manifests itself in the low-temperature ($T \ll J$) behavior of thermodynamic quantities. 

In the thermodynamic limit, the free energy of the considered quantum spin chain is given by
\begin{equation}
F=-T\sum_{k}\ln\left(2\cosh\frac{E(k)}{2T}\right)=-T\int_{-W}^{W}d\omega\rho(\omega)\ln\left[2\cosh(\omega/2T)\right]\, ,
\end{equation}
where the bandwidth is $W=J+J^{\ast}/4$. Accordingly, the heat capacity is
\begin{equation}
C=\frac{1}{4T^{2}}\sum_{k}
\frac{E^{2}(k)}{\cosh\left(\frac{E(k)}{2T}\right)}=
\frac{1}{4T^{2}}\int_{-W}^{W}d\omega\rho(\omega)
\frac{\omega^2}{\cosh\left(\frac{\omega}{2T}\right)}
\end{equation}

In the limiting case $\alpha=\alpha_{c}$, where $\rho(\omega)\simeq \omega^{-2/3}$, we obtain 
\begin{equation}
C\sim \frac{1}{\left(J^{\ast}_{c}\right)^{1/3}T^{2}}\int_{0}^{T}d\omega\omega^{4/3} \sim (T/J^{\ast}_{c})^{1/3}\, .
\end{equation}
\begin{figure}[!t]
\begin{center}
 \includegraphics[width =0.7\columnwidth]{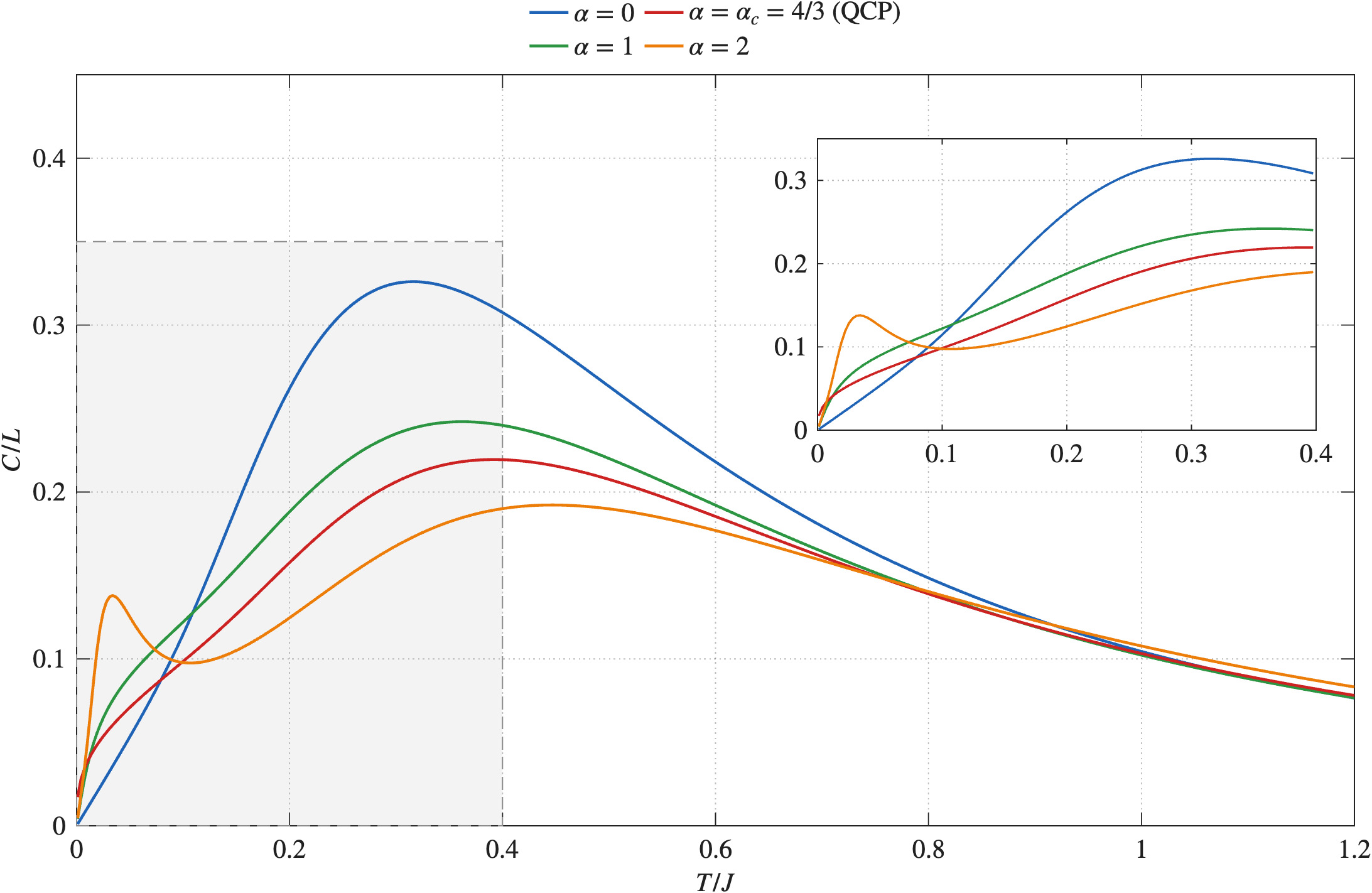}
  \caption{Heat capacity of the system as a function of temperature for $\alpha=0$ (blue line), $\alpha=1$ (green line), $\alpha=2$ (orange line), and at the QCP $\alpha=4/3$ 
  (red line). The inset shows the low-temperature region $0<T/J<0.4$, where the nonlinear behavior of the heat capacity at $\alpha=4/3$ and in the vicinity of the critical point ($\alpha=1$) is clearly seen. The sharp maximum of the heat capacity for 
  $\alpha=2$ at $T\simeq 0.04J$ results from the dominant square-root-type singularity in the density of states near the appeared additional local extrema.}
  \label{fig:Heatcapacity}
 \end{center} 
\end{figure}

In Fig.~\ref{fig:Heatcapacity}, we plot the heat capacity of the system as a function of the dimensionless parameter $T/J$ at $\alpha=0$ (blue line), $\alpha=1$ (green line), $\alpha=2$ (orange line), and at the QCP $\alpha=4/3$ (red line). The inset shows the low-temperature region $0<T/J<0.4$, where the nonlinear behavior of the heat capacity at $\alpha=4/3$ as well as in the vicinity of the critical point ($\alpha=1$) is clearly seen. The sharp maximum of the heat capacity at $\alpha=2$ results from the dominant square-root-type singularity in the density of states near the appearence of additional local extrema.
\subsection{Correlation functions}
\label{sec:GSPD}

The QPT at $\alpha=\alpha_{c}$ also manifests itself in the large-distance behavior of the transverse correlation function. Corresponding changes are perfectly seen already via behavior of the longitudinal spin-spin correlation functions 
\bea
&K^{z}(r)  =  \langle S_{n}^{z}S_{n+r}^{z} \rangle\, ,&
\eea
The spin correlations functions for the ordinary $XX$ chain 
($\alpha=0$) were studied in the classical paper by Lieb, Schulz, and
Mattis \,\cite{LSM_61}. In our calculations of the correlation functions, we follow the approach used in Ref. \cite{LSM_61}. In this approach, the correlation functions are given in terms of Toeplitz determinants. Using the Jordan-Wigner transformation (\ref{JWtransformation}), one can easily find that
\bea\label{g}
&{\cal K}^{z}(r) = det \, \hat{{\bf g}}=\frac{1}{4}
\left| \begin{array}{cc} 
g(0) & g(r) \\
g(r) & g(0) 
\end{array} \right| = \frac{1}{4} \left(g^{2}(0)-g^{2}(r) \right),&
\eea
where
\be\label{gr1}
g(r)=\frac{2}{\pi r }~\sin(\pi r/2) - \delta_{0r}
\ee
for $\alpha<\alpha_{c}$, and 
\be\label{gr2}
g(r)=\frac{2}{\pi r} \left[\sin(\frac{\pi}{2}-k_{0})r -\sin(\frac{\pi}{2}r)+
\sin(\frac{\pi}{2}+k_{0})r\right] = 
\frac{2}{\pi r} \sin(\frac{\pi}{2}r)\left[2\cos(k_{0}r)-1\right]
\ee
for $\alpha > \alpha_{c}$.

After straightforward calculations, we obtain, for $\alpha < \alpha_{c}$,
\be\label{zzCorreelationAlphaless}
{\cal K}^{z}(r)=  m_{z}^{2}-\frac{1}{\pi^{2}}\frac{\sin^{2}(\frac{\pi}{2}r )}{r^{2}}
\ee
and, for $\alpha > \alpha_{c}$,
\be\label{zzCorreelationAlphamore}
{\cal K}^{z}(r)=  m_{z}^{2}- \frac{1}{\pi^{2}}\frac{\sin^{2}(\frac{\pi}{2}r )}{r^{2}}\left[2\cos(k_{0}r)-1\right]^{2}\, .
\ee
As seen, the quadratic decay of the longitudinal correlations remains 
unchanged across the transition. However, additional oscillations in the longitudinal correlation function, $\sim\cos(2k_{0}r)$ and $\sim\cos(k_{0}r)$, 
associated with the additional Fermi points are clearly observed.

Below, the gapless phase with the power-law decay of spin-spin correlations 
given by Eqs. (\ref{zzCorreelationAlphaless}) is called the {\em Spin Liquid I} (SL-I) phase, while the gapless phase with the power-law decay of spin-spin correlations given by Eqs. (\ref{zzCorreelationAlphamore}) is called the {\em Spin Liquid II} (SL-II) phase. The behavior of the correlation functions given by Eqs. 
(\ref{zzCorreelationAlphaless})--(\ref{zzCorreelationAlphamore})
is fully consistent with the standard conformal-field-theory results for 
systems with several gapless excitations \cite{Frahm_92,ZKZ_2001}.

\section{Magnetic properties at $T=0$.}
\label{sec:magnet}

Besides the thermodynamic characteristics, peculiarity of the spectral properties of the system at the QPT  point manifests itself also in the magnetic properties of the system both at $T=0$ and finite $T$. In the presence of external magnetic field the Hamiltonian is given by 
\bea \label{GXYhamiltonian_MagField}
 H_{XzzX}  &=&   -J\sum_{n=1}^{N}\left(S^{x}_{n}S^{x}_{n+1} + 
S^{y}_{n}S^{y}_{n+1}\right)\nonumber\\ 
&&- J^{\ast} \sum_{n=1}^{N}  \left(S^{x}_{n}S^{x}_{n+3} +
S^{y}_{n}S^{y}_{n+3}\right)S^{z}_{n+1}S^{z}_{n+2} -\sum_{n=1}^{N}H_{n}S^{z}_{n}\, . 
\eea
Below we consider the case, where $H_{n}=H+(-1)^{n}H_{st}$.

\subsection{Uniform external magnetic field}

Let us first consider the case of a uniform magnetic field. In this case, at $T=0$, the magnetization per site is determined by the density of SFs in the ground state:
\bea\label{Magnetization_uniform}
& m^{z}(H)=\frac{1}{L}\sum_{n}\langle S_{n}^{z}\rangle
=\frac{k^{+}_{F}(H)+k^{-}_{F}(H)}{\pi}-\frac{1}{2}\, . &
\eea 
Here, the Fermi momenta $k^{\pm}_{F}(H)$ are solutions of the equation
\be\label{eqs:for_kF_H}
\alpha\cos^{3}k+(1-3\alpha/4)\cos k +H/J=0\, .
\ee 
Although analytic expressions for the solutions of the cubic equation (\ref{eqs:for_kF_H}) are known, we do not present them here and restrict ourselves to a qualitative discussion of the magnetization and magnetic susceptibility based on a numerical solution of Eq. (\ref{eqs:for_kF_H}). An exception is the quantum critical point $\alpha=\alpha_{c}$, where the solution of Eq. (\ref{eqs:for_kF_H}) takes the compact form
\bea\label{KF_H}
&k_{F}(H)=\arccos\left(-H/J^{\ast}\right)^{1/3}\, .&
\eea 
Therefore, at the QCP, the magnetization per site is
\bea\label{Magnetization_QCP}
& m^{z}(H)=\frac{1}{\pi}\arccos\left(-H/J^{\ast}\right)^{1/3}
-\frac{1}{2}
\sim \left(H/J^{\ast}\right)^{1/3}\, ,
\qquad H \ll J^{\ast}\, . &
\eea 
Accordingly, the magnetic susceptibility is given by
\be
\chi(H)=\frac{\partial m^{z}(H)}{\partial H}
=\frac{2}{\pi J^{\ast}}\left(H/J^{\ast}\right)^{-2/3}
\frac{1}{\sqrt{1-(H/J^{\ast})^{2/3}}}
\sim \frac{2}{\pi J^{\ast}}
\left(H/J^{\ast}\right)^{-2/3}\, ,
\qquad H \ll J^{\ast}\, .
\ee
Thus, at the QCP, the magnetic susceptibility follows the density of states and diverges as $\left(H/J^{\ast}\right)^{-2/3}$.

\begin{figure}[!t]
\begin{center}
 \includegraphics[width =0.85\columnwidth]{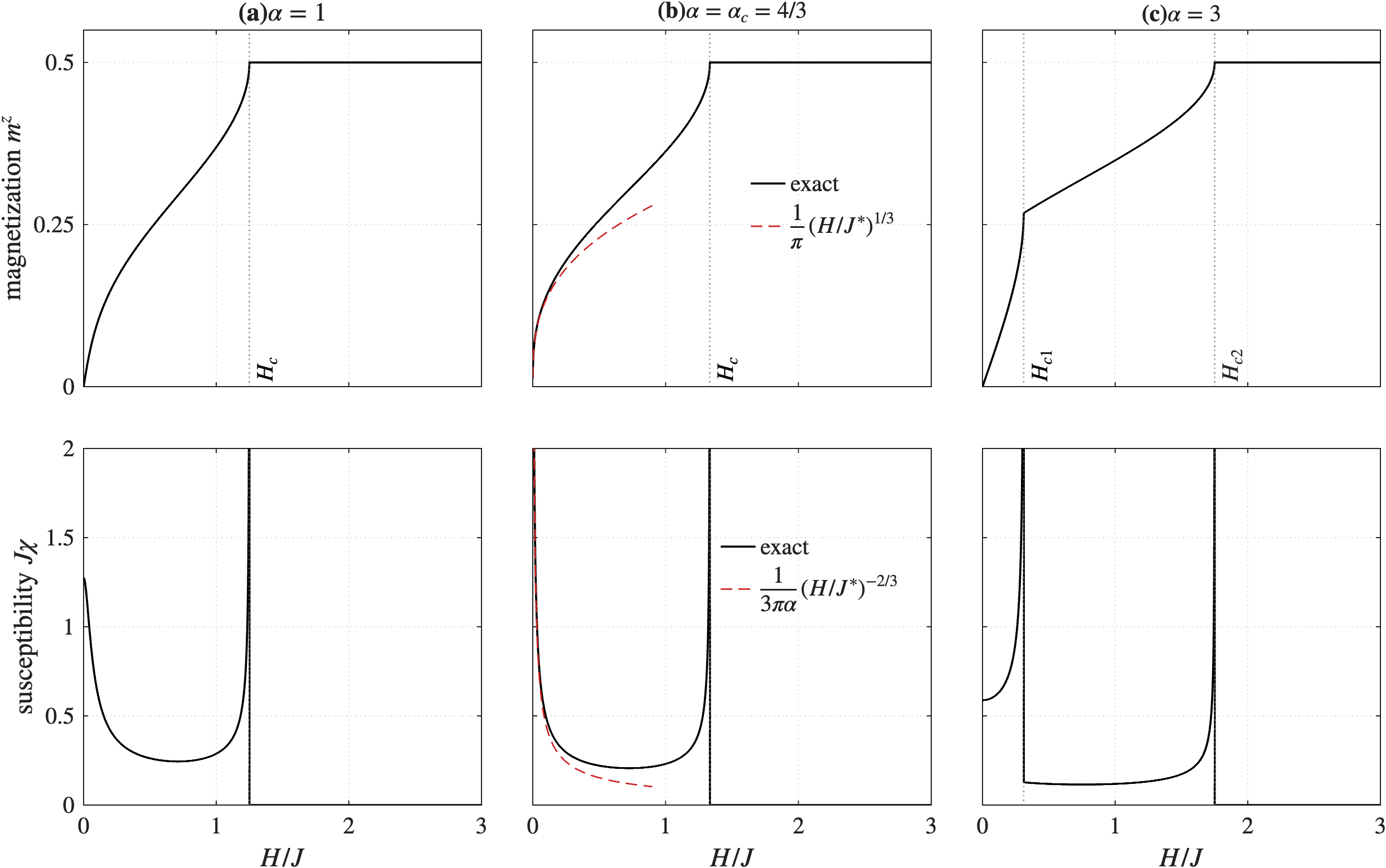}

 \caption{Magnetization and magnetic susceptibility as functions of the uniform magnetic field at $\alpha=1$ (a), $\alpha=4/3$ (b), and $\alpha=3$ (c). The dashed lines mark the singular characteristics of the magnetization and magnetic susceptibility at the critical point.}
 \label{fig:M_and_Chi}
\end{center} 
\end{figure}

For $\alpha \neq \alpha_{c}$, the magnetic properties of the system depend on $\alpha$. In Fig.~\ref{fig:M_and_Chi}, we plot the magnetization and magnetic susceptibility as functions of the dimensionless magnetic field $h=H/J$ at $\alpha=1$, $\alpha=4/3$, and $\alpha=3$. For $\alpha < \alpha_{c}$ and $H<H_{c}=\pm W$, the system is characterized by two Fermi points, and the density of states near these points remains finite. As a result, the magnetization increases monotonically with increasing magnetic field, while the magnetic susceptibility remains finite, although it becomes large as $\alpha$ approaches $\alpha_{c}$, until the magnetic field reaches the band edge at $H=H_{c}$. Near the transition into the fully polarized phase, the density of states exhibits the standard square-root singularity, and the magnetic susceptibility diverges as
\be\label{eq:chi-Hc}
\chi(H) \sim (H_{c}-H)^{-1/2}\, . 
\ee

\begin{figure}[!t]
\begin{center}
  \includegraphics[width =0.7\columnwidth]{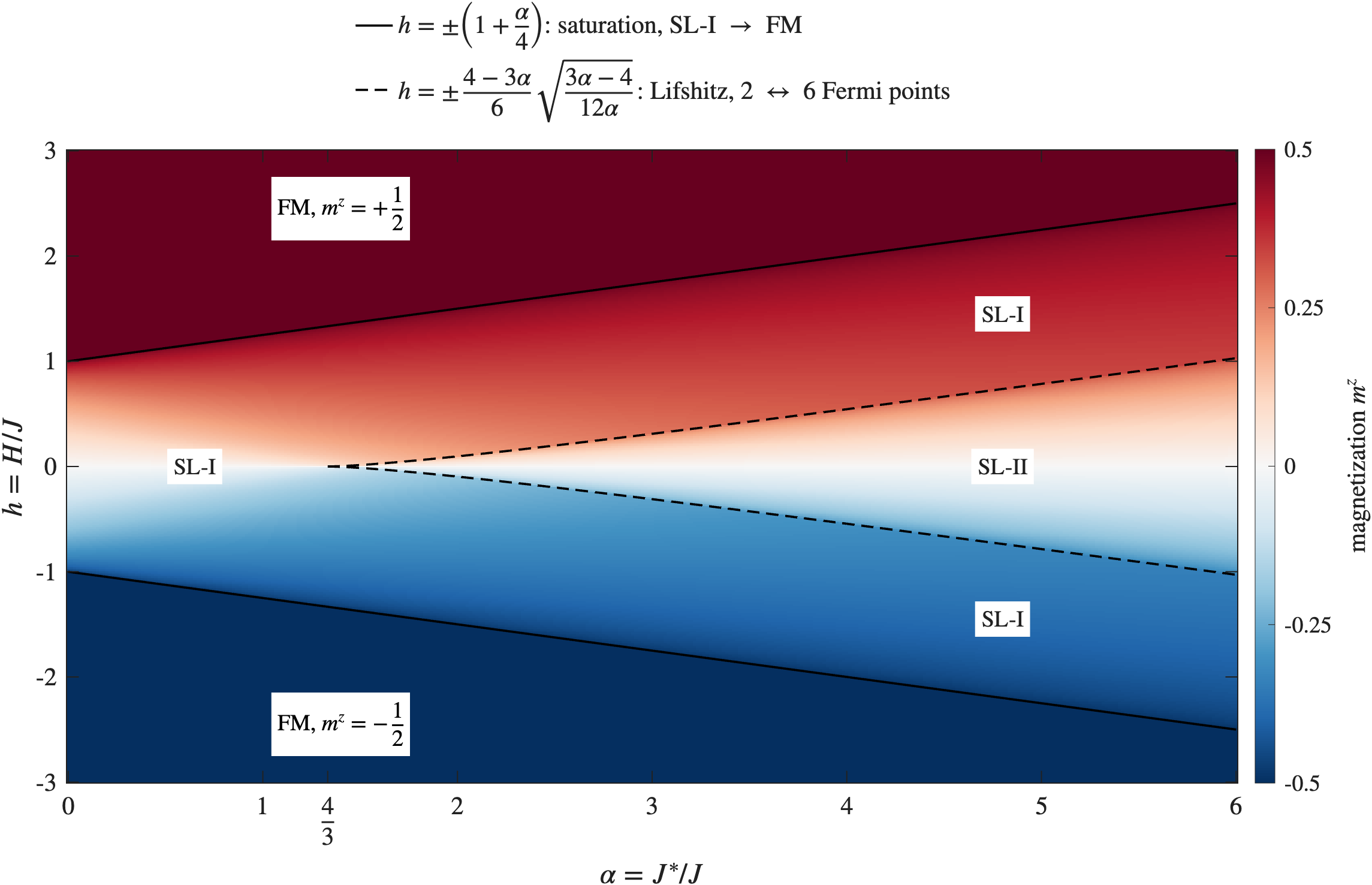}  \caption{Magnetic phase diagram of the model as a function of the magnetic field and the parameter $\alpha$. The solid lines $h=\pm(1+\alpha/4)$ mark transitions into the fully polarized ferromagnetic state, while the dashed lines $h=\pm\frac{4-3\alpha}{6}\sqrt{\frac{3\alpha-4}{12\alpha}}$ mark, with increasing $\alpha$, the topological Lifshitz transition from the phase with two Fermi points to the phase with six Fermi points.}
  \label{fig:Magnetic_PD_Uni_field}
\end{center} 
\end{figure}

For $\alpha > \alpha_{c}$, there are six Fermi points, and the spectrum exhibits non-monotonic behavior. Therefore, with increasing magnetic field, the magnetization shows more complex behavior and develops an additional singularity when the field reaches the top of the bump in the dispersion curve at
\begin{equation} \label{CriticalField_Hc1}
H=H_{c1}= \frac{J^{\ast}}{4}\left(1-\frac{4}{3\alpha}\right)^{3/2}\, .
\end{equation} 
Near this local maximum, the density of states again exhibits a square-root singularity, and the magnetic susceptibility diverges as $\chi(H)\sim (H_{c1}-H)^{-1/2}$. With further increase of the magnetic field, the system again has two Fermi points, the magnetization increases almost linearly, and the susceptibility remains nearly constant up to the upper critical field. Near the transition into the fully polarized phase, the magnetization again shows standard square-root behavior, while the magnetic susceptibility exhibits the singularity given by (\ref{eq:chi-Hc}). The magnetic phase diagram of the system as a function of the uniform magnetic field $h$ and the parameter $\alpha$ is summarized in Fig.~\ref{fig:Magnetic_PD_Uni_field}.

\subsection{Alternating magnetic field}

\begin{figure}[!t]
\begin{center}
  \includegraphics[width =0.4\columnwidth]{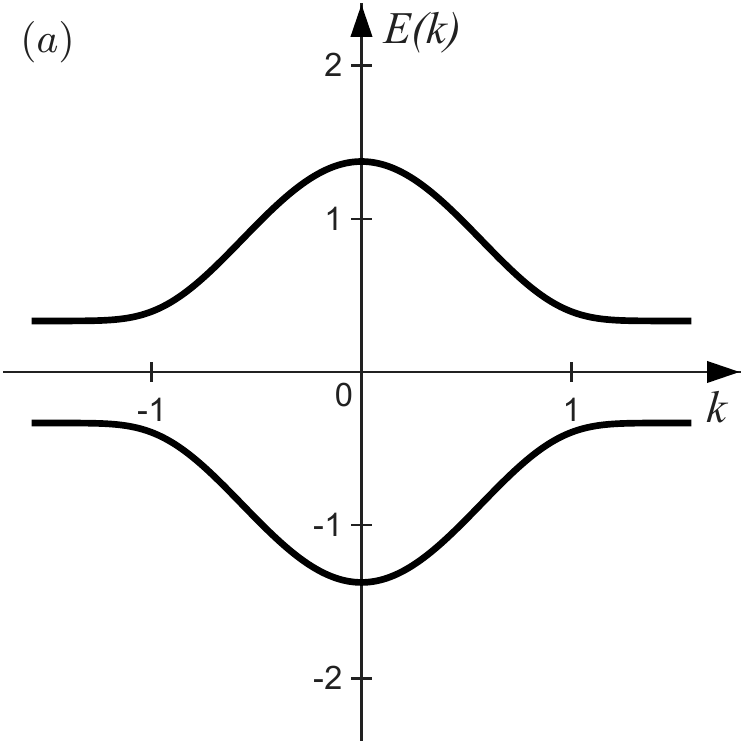}
   \includegraphics[width =0.4\columnwidth]{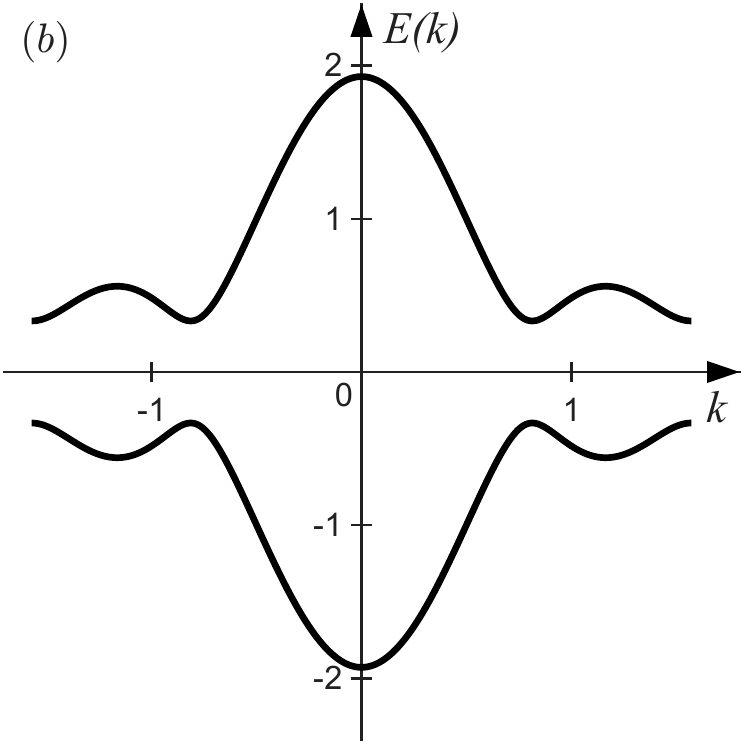}
  \caption{ Dispersion relation for equivalent spinless fermion model in the presence of staggered magnetic field $H_{1}=0.5J$ at the QPT point $\alpha=\alpha_{c}=4/3$ (a) and at  $\alpha=3.6$ (b). On the left Figure the wide, almost flat segment of the dispersion relations in the vicinity of Brillouin zone borders $k=\pm \pi/2$   is perfectly seen.}
\label{fig:Spectrum_Stag_Field}
 \end{center} 
\end{figure}

In the presence of an alternating magnetic field $H=H_{0}+(-1)^n H_{1}$, diagonalization of the corresponding spinless fermion Hamiltonian is straightforward (see Ref. \cite{Franchini_2017}), and the spectrum of 
the system is given by
\begin{eqnarray}\label{Spectrum_alt_MF}
&E(k)=H_{0}\pm \sqrt{ \left[J\left(1-\frac{3\alpha}{4}\right)\cos k+J^{\ast}\cos^{3} k\right]^{2}+H_{1}^{2}}  \qquad \left\{-\pi/2\leq k<\pi/2\right\}&. 
\end{eqnarray}
At the QPT point and in the absence of the uniform component of the magnetic field, 
the system is characterized by strong flattening of the dispersion relation near the new Brillouin-zone boundaries at $k=\pm \pi/2$ (see Fig.~\ref{fig:Spectrum_Stag_Field}). At the critical point $\alpha=4/3$ and near the zone boundaries $k=\pm (\pi/2-p)$, with $p \ll \pi/2$, the spectrum is given by
\begin{eqnarray}\label{Spectrum_alt_MF_CP}
E(k)\simeq H_{0}\pm H_{1}\sqrt{1+\left(J^{\ast}/H_{1}\right)^2 p^{6}}
\simeq H_{0}\pm H_{1}\pm Cp^{6}+{\cal O}\left(p^{8}\right),
\end{eqnarray}
where $C=J^{\ast\,2}/H_{1}$.

For $|H_{0}|<H_{c}=H_{1}$, the excitation spectrum is gapped, and the ground state of the equivalent SF system exhibits a long-range-ordered (LRO) charge-density wave. In terms of the original spin chain, this corresponds to an LRO N\'eel antiferromagnetic state controlled by the staggered component of the magnetic field. However, for $|H_{0}|>H_{c}$, a transition to a metallic (i.e., spin-liquid) phase occurs.  At the Lifshitz transition point, owing to the extremely flat spectrum, the density of states behaves as $\rho(\omega)\sim \omega^{-5/6}$. As a result, the net magnetization of the system increases as
\begin{equation}
m =\int_{0}^{H-H_{c}}\rho(\omega)d\omega \sim (H-H_{c})^{1/6}
\end{equation}
and the magnetic susceptibility diverges as
\begin{equation}
\chi(h)\simeq (H-H_{c})^{-5/6}\, .
\end{equation}
Away from the critical point, at $\alpha \neq \alpha_{c}$ the gap in the dispersion relation opens again at the Fermi points, at $\alpha<\alpha_{c}$ only at the borders of the Brillouin zone  $\pm k_{F}=\pi/2$, while at $\alpha > \alpha_{c}$ at all Fermi points given by Eq.(\ref{FermiPointsAlpha_1}) (see Fig. \ref{fig:Spectrum_Stag_Field}b.) The density of states is characterized by the singularity in the vicinity of the minimums of the dispersion relation, but of the square-root type $\rho(\omega)\sim \omega^{-1/2}$. Consequently the magnetic phase transition from LRO AFM  phase at $|H_{0}|<H_{c}$ into a paramagnetic spin-liquid phase at $|H_{0}|>H_{c}$ is accompanied by the emergence of  magnetization in a square-root form $m \sim (H-H_{c})^{1/2}$ and therefore  the magnetic susceptibility diverges as $\chi(h)\simeq (H-H_{c})^{-1/2} $.

\section{Conclusion}
\label{sec:Summary}

To summarize, we have studied the ground-state magnetic phase diagram and low-temperature thermodynamics of the spin $S=1/2$ extended $XzzX$ chain with four-spin interaction. We have exploited the Jordan-Wigner transformation and shown that, in terms of the equivalent spinless-fermion representation, the model describes free lattice fermions with nearest-neighbor ($J$) and next-next-next-neighbor ($J^{\ast}/4$) hopping. The spinless-fermion dispersion relation exhibits the change in Fermi-surface topology characteristic of a Lifshitz transition at $J^{\ast}>J^{\ast}_c=4J/3$, where the number of Fermi points triples. The critical point marks the transition between two gapless spin-liquid phases characterized by different behaviors of the spin-spin correlation functions.

At the transition point, the dispersion relation near the Fermi points $k_{F}=\pm \pi/2$ exhibits strong flattening, $E(\pm\pi/2+p)\sim p^{3}$. As a result, the system exhibits unusual singular properties. Namely, the density of states is singular, $\rho(\omega)\sim(\omega/J)^{-2/3}$, the heat capacity shows the unconventional temperature dependence $C\sim(T/J^{\ast})^{1/3}$, and the magnetic susceptibility diverges as $\chi(H)\sim(H/J^{\ast})^{-2/3}$.

The band-flattening effect is even stronger when the magnetic properties of the system are considered at the critical point $J^{\ast}_c=4J/3$ in the presence of an alternating magnetic field $H(n)=H_{0}+(-1)^{n}H_{1}$. In this case, the system is characterized by a rich ground-state phase diagram including fully polarized ferromagnetic, gapped antiferromagnetic (AFM), and partially polarized spin-liquid phases. Just above the transition from the gapped AFM phase ($H<H_{c}=H_1$) to the gapless polarized spin-liquid phase ($H>H_{c}$), the dispersion relation near the Brillouin-zone boundaries $\pm\pi/2$ is given by $E(\pm\pi/2+p)\simeq \pm Cp^{6}$. As a result, the system exhibits a rapid increase in magnetization, $m\sim(H-H_c)^{1/6}$, and singular magnetic susceptibility, $\chi(H)\sim(H-H_c)^{-5/6}$. Accordingly, the heat capacity at this point behaves as $C(H_{c})\sim(T/J^{\ast})^{1/6}$.

In conclusion, we have shown that the spin $S=1/2$ extended $XzzX$ chain with four-spin interaction exhibits strong band flattening at the critical point associated with the topological Lifshitz transition, leading to several unusual magnetic and thermodynamic properties. Recent impressive progress in developing schemes for the study of lattice transport phenomena in the case of single-particle Hamiltonians, based on the controlled momentum-space dynamics of ultracold atomic matter waves \cite{Gadway_2015,Gadway_etal_2016,Gadway_etal_2018} makes it very promising to check  the obtained results experimentally,  using the optically trapped ultracold atomic systems.

\section{Acknowledgments}

The authors would like to thank George Jackeli and Alexander A. Nersesyan for stimulating interest in this work with numerous enlightening discussions.\\

\vspace{5mm}

{\bf Author contributions}

All authors contributed equally to the paper. All authors have read and approved the final manuscript. All authors contributed equally to the analytical calculations contained in the present manuscript.
\medskip

{\bf Data availability statement} 

This manuscript does not have associated data or the data will not be deposited. 
\medskip

{\bf Declarations}

{\sf Conflict of interest} The authors have no competing interests.

\end{document}